\documentclass[preprint,review,12pt]{elsarticle}

\usepackage{amssymb}

\usepackage{lineno}

\usepackage{xcolor}
\usepackage{ccaption}

\usepackage{float}
\usepackage[numbers]{natbib}
\biboptions{comma,sort&compress}
\usepackage{amsthm}
\usepackage{mathrsfs,amsmath}
\usepackage{amsbsy}
\usepackage{graphicx}
\usepackage{subcaption}
\usepackage{siunitx}
\usepackage[english]{babel}
\usepackage{supertabular}
\usepackage{multirow}
\usepackage{bm} 
\usepackage[nameinlink,noabbrev]{cleveref}
\usepackage{array}
\usepackage{amssymb}
\usepackage{lineno}
\usepackage{longtable}
\usepackage{booktabs}
\usepackage[normalem]{ulem}
\usepackage{amsmath}

\graphicspath{{images/}}

\journal{ }

\begin{document}

\begin{frontmatter}

\title{\textcolor{black}{ Identifying regions of importance in wall-bounded turbulence through explainable deep learning}}

\author[KTH,CMT]{A. Cremades}
\ead{andrescb@kth.se}
\author[IUMPA]{S. Hoyas}
\author[MEL]{\textcolor{black}{ R. Deshpande}}
\author[CMT]{P. Quintero}
\author[UEd2]{M. Lellep}
\author[MEL]{\textcolor{black}{ J. Lee}}
\author[MEL]{\textcolor{black}{ J. P. Monty}}
\author[MEL]{\textcolor{black}{ N. Hutchins}}
\author[UEd]{M. Linkmann}
\author[MEL]{\textcolor{black}{ I. Marusic}}
\author[KTH]{R. Vinuesa}
\ead{rvinuesa@mech.kth.se}

\address[KTH]{FLOW, Engineering Mechanics, KTH Royal Institute of Technology, SE-100 44 Stockholm, Sweden}
\address[CMT]{CMT-Motores Térmicos, Universitat Politècnica de València, Camino de Vera s/n, Valencia, 46022, Spain}
\address[IUMPA]{Instituto Universitario de Matemática Pura y Aplicada, Universitat Politècnica de València, 46022 Valencia, Spain}
\address[MEL]{\textcolor{black}{ Department of Mechanical Engineering, University of Melbourne, Parkville, VIC 3010, Australia}}

\address[UEd2]{SUPA, School of Physics and Astronomy, The University of Edinburgh, James Clerk Maxwell Building, Peter Guthrie Tait Road, Edinburgh EH9 3FD, UK}

\address[UEd]{School of Mathematics and Maxwell Institute for Mathematical Sciences, University of Edinburgh,
Edinburgh EH9 3FD, UK}

\begin{abstract}

Despite its great scientific and technological importance, wall-bounded turbulence is an unresolved problem \textcolor{black}{ in classical physics} that requires new perspectives to be tackled. One of the key strategies has been to study interactions among the \textcolor{black}{ energy-containing} coherent structures in the flow. Such interactions are explored in this study for the first time using an explainable deep-learning method. The instantaneous velocity field \textcolor{black}{ obtained from a turbulent channel flow simulation} is used to predict the velocity field in time through a \textcolor{black}{ U-net architecture}.  Based on the predicted flow, we assess the importance of each structure for this prediction using the game-theoretic algorithm of SHapley Additive exPlanations (SHAP). This work provides results in agreement with previous observations in the literature and extends them by \textcolor{black}{ revealing that the most important structures in the flow are not necessarily the ones with the highest contribution to the Reynolds shear stress. We also apply the method to an experimental database, where we can identify completely new structures based on their importance score. This framework has the potential to shed light on numerous fundamental phenomena of wall-bounded turbulence, including novel strategies for flow control.}
\end{abstract}


\end{frontmatter}

\section*{Introduction} \label{sec:introduccion}

Approximately 140 years ago, Osborne Reynolds published the first and most influential scientific article on turbulent flows \cite{Rey83}. One of the main conclusions of this study was the fact that the Navier--Stokes equations, which describe the behavior of any flow, can only be solved analytically for elementary flow configurations. For nearly a century, the study of turbulence has relied on experimental measurements~\cite{tay38,kli67,tow76} and theoretical considerations~\cite{kol41a}. Almost all flows of practical interest are turbulent, except those relevant to lubrication~\cite{pop00}. In fact, one of the most crucial challenges nowadays, namely the current climate emergency, is closely connected with turbulence and a better understanding of the dynamics of turbulent flows is necessary to reduce greenhouse-gas \textcolor{black}{ emissions.} Approximately 30\% of the energy consumption worldwide is used for transportation~\cite{IEA2020}, which, due to the increase in drag caused by turbulent flow, is a problem very closely connected with wall-bounded turbulence. Furthermore, turbulence is critical in combustion processes~\cite{Kerstein2002,Peters2009} and aerodynamics~\cite{Panagiotou2020,Vinuesa2022_2}. It is also essential in energy generation \cite{Lubitz2014,Optis2019} or urban pollution \cite{Ulke2001,Solazzo2008}, to name just a few. Indeed, some estimations indicate that 15\% of the energy consumed worldwide is spent near the boundaries of vehicles and is therefore related to turbulent effects \cite{Jimenez2013}. 

The main challenge is the fact that turbulence is a multi-scale phenomenon in both time and space. The energy is mainly transferred from the largest to the smallest scales of the flow, where it is dissipated \cite{kol41a}, although there is also an energy path in the opposite direction~\citep{Cardesa_science}. There are several orders of magnitude between these scales for any flow in engineering, \textcolor{black}{ and this scale separation increases at progressively higher Reynolds numbers}. In the presence of a wall, this energy cascade is even more complicated due to the energy and momentum transfer from the wall to the outer flow~\cite{Jimenez2018}. This multi-scale behavior implies that integrating numerically the Navier--Stokes equations requires extremely fine computational meshes, leading to a prohibitive computational effort for practical applications.  
 
In the 1980s, supercomputers became powerful enough to integrate these equations numerically in some canonical geometries. \citet{kim87} simulated the simplest complete example of a wall-bounded flow, {\it i.e.,} a turbulent channel. They performed a direct numerical simulation (DNS), where all the spatial and temporal scales of the flow are resolved. Note that in DNS there are no additional hypotheses beyond the fact that the flow is governed by the Navier--Stokes equations. This numerical technique provides a complete flow characterization, and almost any imaginable quantity can be computed. Thus, DNS can provide a large amount of high-quality data, and simulations in the Petabyte scale are becoming progressively more common~\cite{hoy22}. This enables fully characterizing the kinematics of wall-bounded turbulent flows. However, describing the dynamics of these flows is still an open challenge. It is then essential to develop novel methods to solve the questions posed 140 years ago.  

One of the most successful ideas for studying turbulent flows focuses on the relationship among the different scales and coherent structures of the flow \cite{cardesa2017turbulent,Jimenez2018}. Note that different definitions of coherent structure have been proposed in the literature. The first examples of coherent structure are the streamwise streaks~\cite{kli67} and the Reynolds-stress \textcolor{black}{ events}~\cite{lu73}, which were first observed experimentally. The latter, also called \textcolor{black}{ Q events,} are the object of our work. Coherent Q structures are flow regions associated with momentum transfer and turbulent-kinetic-energy (TKE) production. Two particular Q events defined below, ejections and sweeps, are the main contributors to the exchange of streamwise momentum. This process is the main energy source for all the structures present in turbulent flows~\cite{wal72, lu73}. \textcolor{black}{ Note that the Q events} are also responsible for the generation of turbulent drag. Even with extensive studies on the contribution of the various coherent structures to the dynamics of turbulent flows, a clear understanding of their actual role still needs to be provided~\cite{Jimenez2018}.

 This study proposes a new technique for the study of wall-bounded turbulence. We have developed a novel methodology based on explainable artificial intelligence (XAI) to gain a more profound knowledge of the flow physics and to evaluate the contributions of the Q events to flow-field prediction. The methodology \textcolor{black}{ is based on a particular type of deep convolutional neural network (CNN)}~\cite{lecun2015}, \textcolor{black}{ namely the U-net~\cite{ronneberger2015},} and the Shapley additive explanation (SHAP) values~\cite{lundberg2017,Meng2020,LunChau2022}. CNNs can effectively extract the spatial information in the flow data~\cite{guastoni2021}, both in two and three dimensions. The SHAP algorithm is a game-theoretic method that calculates the importance of each input feature on the \textcolor{black}{ U-net} prediction. SHAP has been shown to correctly identify key aspects of the near-wall cycle that sustains turbulence close to onset~\cite{Lellep2022}. Thus, the main novelty of this work is the explainability of fully-developed turbulence through artificial intelligence. We calculate the relative importance of each Q event for the \textcolor{black}{ U-net} prediction through SHAP. In doing so, we identify, in a purely data-driven method (without any hypothesis about the physics of the flow), relevant physical processes governing the dynamics of wall-bounded turbulence. 
  
\begin{figure}[H]
\centering
\includegraphics[width=0.95\textwidth]{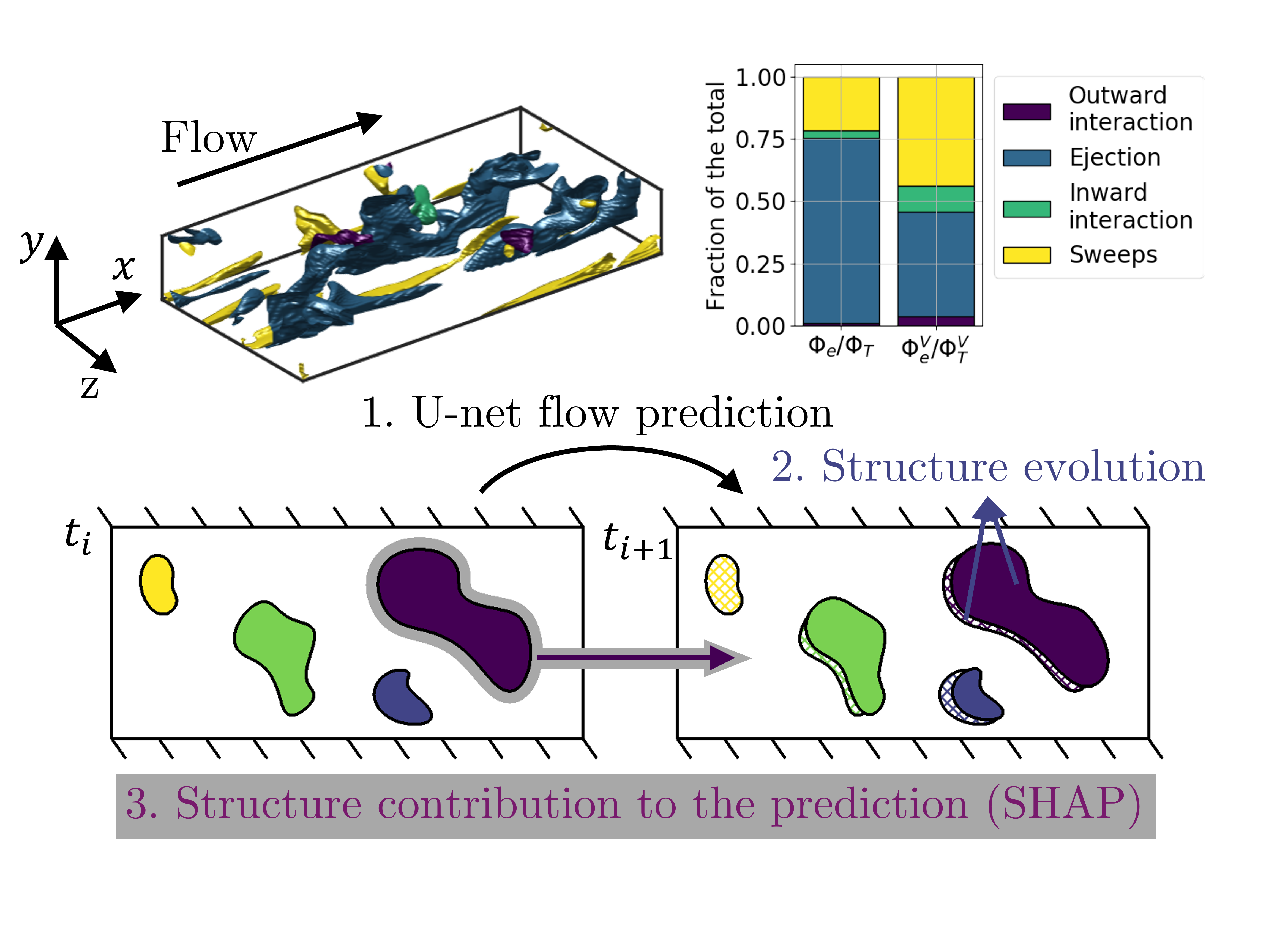}
\caption{{{\bf Conceptual map of the workflow employed in this study.}} \textcolor{black}{(Top-left) Instantaneous Reynolds-stress (Q) events identified in a turbulent channel. Four different kinds of structures exist based on the quadrant analysis \cite{Lozano2014}: outward interactions (purple), ejections (blue), inward interactions (green) and sweeps (yellow). (Top-right) Total contribution, $\Phi_e / \Phi_T$, (left column)  and total contribution per unit volume, $\Phi_e^v / \Phi_T^v$, (right column) of each event type to the U-net prediction. Their definition and implications are discussed in the Results section. (Bottom) Workflow comprising three steps: 1) a U-net is used to predict the next instantaneous flow field (time $t_{i+1}$) based on the current one ($t_i$); 2) the structures evolve, so some may dissipate in the next field (yellow), others may be convected (rest of colors), and some may even merge into larger ones (not shown); 3) calculation of the contribution of each structure (gray shade) to the prediction of the next field. The error on the prediction of the flow field of the U-net in $t_i$ with respect to the simulated flow in $t_{i+1}$ is used to determine the importance of every single structure.  In this way, it is possible to rank the various structures in terms of their relative importance to predict the next instantaneous field. The workflow is performed on the full three-dimensional data but shown on a vertical slice of the turbulent channel here for simplicity.}}
\label{fig:conceptual_map}
\end{figure}

To accomplish this objective, we will first show how \textcolor{black}{ U-nets} can predict the evolution of turbulent channel flow, \textcolor{black}{ extending our} earlier work~\cite{Schmekel2022}. We start with a database of \textcolor{black}{ 6,000} instantaneous realizations \textcolor{black}{ obtained from turbulent channel flow simulations,} \textcolor{black}{ see the Methods section for additional details on the data generation.} For every field, the domain is segmented into Q events (see \textcolor{black}{ the Results} section), and each one of these structures is considered an input feature to the SHAP algorithm. SHAP ranks the importance of each structure for predicting the following flow field, as shown schematically in Figure~\ref{fig:conceptual_map}. This workflow consists of three main stages: prediction of the flow through a \textcolor{black}{ U-net}, determination of the structure evolution (advance a time step in the simulation), and quantification of the importance of each coherent turbulent structure using SHAP values (and SHAP values per unit of volume) comparing the predicted solution with the simulated flow field in the next time step. By analyzing the characteristics of the highest-ranked structures, we can shed light on the dynamics of wall-bounded turbulence, with direct implications on the questions described above. We find coherent structures representing ejections, where \textcolor{black}{ flow regions} with low streamwise velocity move from the near-wall towards the outer region; and sweeps, where \textcolor{black}{ the flow regions} with high streamwise velocity move from the outer region towards the wall. \textcolor{black}{ Our} study confirms \textcolor{black}{ and extends} the results obtained by other authors~\cite{Lozano2012,Jimenez2018}, introducing the usage of XAI to \textcolor{black}{ define an objective metric to identify the most important coherent structures in the flow. This is not a new question in the study of turbulence: Encinar and Jim\'enez~\cite{encinar_jimenez} studied homogeneous isotropic turbulence by introducing perturbations in the flow and integrating numerically the governing equations to assess their evolution in time. By doing so, they could assess the importance of various types of perturbations, concluding that the strain-dominated vortex clusters are the most important regions of the flow. Their method relies on actively modifying the flow through the introduction of the perturbations and re-simulating it using DNS. An alternative approach to answer this question was proposed by Lozano-Dur\'an and collaborators~\cite{lozano_arranz,lozano_et_al}, who utilize information theory to identify the regions of the flow with the highest impact on future flow states. Their method relies on time series (which can be obtained for instance via modal decomposition), and analyzes the error of one variable when information from another one is removed. This method does not affect the original flow, therefore it is non intrusive, and relies on large amounts of temporal data. The approach adopted in this work  relies on deep learning, and is intrusive on the surrogate model, rather than the original governing equations. The advantage of the present approach is the fact that it can provide a good surrogate of the governing equations even in environments where full access to DNS data is not possible, {\it i.e.} in experiments. This enables assessing the most important regions of the flow at higher Reynolds numbers, where there is a broader hierarchy of dynamically-significant scales~\cite{smits_et_al_2011}, in environments where less data is available. The importance of identifying the most important regions of the flow is manifested {\it e.g.} in flow prediction~\cite{guastoni2021} and flow control~\cite{guastoni_drl}.}

\section*{Results}\label{sec:results}

The geometry of \textcolor{black}{ the} turbulent channel flow comprises two parallel planes at a distance of $2h$, \textcolor{black}{ and a pressure gradient drives the flow in the streamwise direction.} The spatial coordinates are $x$, $y$, and $z$ in the streamwise, wall-normal, and spanwise directions, respectively. The length and width of the channel are $L_x = 2\pi h$, and $ L_z=\pi h$, with streamwise and spanwise periodicity. This computational box is large enough to \textcolor{black}{ adequately represent the one-point statistics} of the flow~\cite{Lozano2014,Lluesma2018}.

 The velocity vector is ${\mathbf U(x,y,z,t)} = (U,V,W)$, where $t$ denotes time. As the flow is fully developed, its statistical information only depends on $y$~\cite{pop00}. Statistically-averaged quantities in $x$, $z$, and $t$ are denoted by an overbar,  whereas fluctuating quantities are denoted by lowercase letters, {\it i.e.}, $U = \overline{U}+u$. Primes are reserved for root-mean-squared (rms) quantities: $u'=\sqrt{\overline{u^2}}$, which constitute a measure of the standard deviation from the mean flow.

The simulation was carried out at a friction Reynolds number $Re_\tau=u_\tau h /\nu = 125$. Note that $\nu$ is the fluid kinematic viscosity and $u_{\tau}=\sqrt{\tau_w/\rho}$ is the friction velocity ($\tau_w$ is the wall-shear stress and $\rho$ the fluid density)~\cite{pop00}, while $Re_\tau$ is the main control parameter. The value of $Re_\tau$ attainable in numerical simulations has been increasing steadily in the last 35 years due to the advances in computational power and numerical methods \citep{kim87,mos99,del04,hoy06,ber14,lee15,yam18,hoy22}. Quantities nondimensionalized with the viscous scales $u_\tau$ and $\nu$ are denoted with a `+' superscript.  Finally, as the channel is statistically symmetric, the upper half-channel statistics are projected symmetrically onto the coordinates of the lower half.

The Q events are coherent regions of instantaneous high Reynolds stress, defined by:
\begin{equation}\label{eq:uv_Qs}
    \vline u(x,y,z,t) v(x,y,z,t) \vline > H u'(y) v'(y). 
\end{equation}

\noindent In this equation, $H$ is the so-called hyperbolic hole, set to a value of $1.75$ \cite{Lozano2012}. Intuitively, equation (\ref{eq:uv_Qs}) identifies the region where the instantaneous value of the Reynolds stress (the left-hand-side) is considerably larger than the product of the standard \textcolor{black}{ deviations of $u$ and $v$.} Q events identify regions that have statistically large magnitudes of Reynolds stress. Based on the classical quadrant analysis~\cite{lu73,Lozano2012}, four types of events can be defined: outward interactions ($u>0$, $v>0$), structures with a high streamwise velocity that move from the wall to the bulk; ejections ($u<0$, $v>0$), structures with a low streamwise velocity which move from the wall to the bulk; inward interactions ($u<0$, $v<0$), structures with low streamwise velocity moving from the bulk to the wall; and sweeps ($u>0$, $v<0$), structures with high streamwise velocity moving from the bulk to the wall, see Figures~\ref{fig:conceptual_map} and \ref{fig:shap_uv}. 

\subsection*{Prediction of the velocity field} \label{sec:pred_flield}

\begin{figure}[H]
\centering
\includegraphics[width=0.85\textwidth]{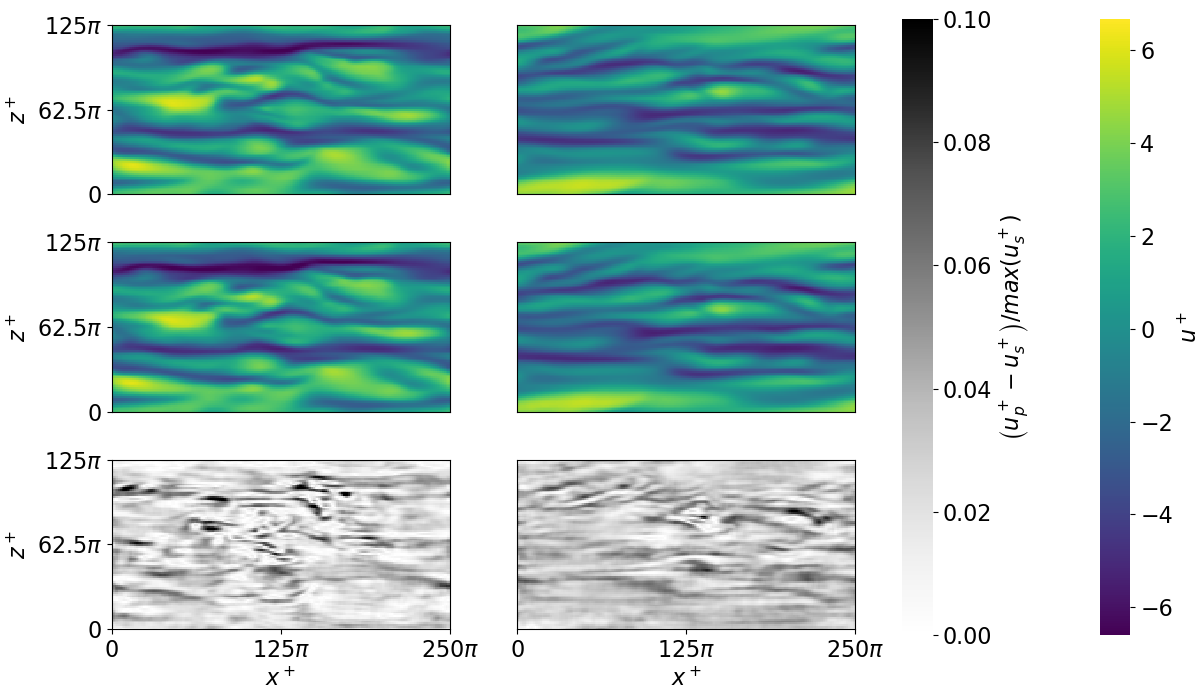}
\caption{\textcolor{black}{\textbf{Comparison of ground truth and prediction for streamwise velocity fluctuations}. We show a representative horizontal slice at $y^+=12$ for a single instantaneous field, where left and right columns represent the lower and upper channel walls. (Top) simulated $u$ velocity field, (middle) predicted velocity field and (bottom) relative error between the two previous fields.  The subscripts $s$ and $p$ correspond to the fields in the reference simulation and the prediction, respectively.}} 
\label{fig:errpred}
\end{figure}

As discussed in the Methods section, a \textcolor{black}{ U-net} is trained to predict \textcolor{black}{ the intantaneous velocity field 5 viscous times into the future, including the three velocity-fluctuation components.} The predictive capabilities of the \textcolor{black}{ U-net} are assessed through the test database, which was not seen by the neural network during training, \textcolor{black}{ and we obtain a relative error of $2\%$ when predicting the three velocity-fluctuation components.} The quality of the prediction can be observed in Figure \ref{fig:errpred}, \textcolor{black}{ where we show a slice of the instantaneous streamwise velocity fluctuation at $y^+=12$ (on both walls) and compare it with the U-net prediction.} The results show \textcolor{black}{ that the U-net architecture constitutes a valid surrogate of the original flow, with a low relative error. Also note} that the employed resolution is sufficient \textcolor{black}{ in this study~\cite{Lellep2022}.}

\subsection*{Explainability of the \textcolor{black}{ U-net} predictions}

The \textcolor{black}{ U-net} is used for calculating the contribution, based on the SHAP values \cite{lundberg2017,Fidel2020,jeyakumar2020}, of the different turbulent structures to the prediction. A deeper explanation of how the SHAP values are used for assessing the importance of each Q event is discussed in the Methods section. The total importance of each typology of Q event, $\Phi_e$, is calculated by summing up the value of every structure belonging to this class of event, $\phi_i^e$, where $i$ indicates a single coherent structure. The total contribution of all events, {\it i.e.} the summation of all the SHAP values, is denoted by $\Phi_T$. Particularizing to a single class $e$ (ejections, sweeps, outward or inward interactions), we can define the percentual importance of this class \textcolor{black}{ as} $\Phi_e / \Phi_T$. \textcolor{black}{ When interpreting SHAP values, the} larger its absolute value, the more critical the structure is to reconstruct the field. In the context of this study, we will use the term {\it importance} to refer to this impact \textcolor{black}{ and we will represent the absolute values of the SHAP scores.} The quantification of $\phi_i^e$ for every structure can be used to evaluate their contribution to the turbulent flow, \textcolor{black}{ and} this magnitude may be evaluated per unit volume. \textcolor{black}{ The} percentual importance per unit of volume is defined as:
\begin{equation} \label{eq:density}
    \Phi_e^v = \sum_{i=1}^{I_e} \left(\frac{\phi_i}{V_i^+}\right)^e, \hspace{1cm} \Phi_T^v = \sum_{e=1}^{4} \Phi_e^v,
\end{equation}
where $I_e$ is the number of structures of type $e$ and $\left(\phi_i/V_i^+\right)^e$ is the SHAP value per unit of volume of the structures type $e$. To avoid spurious results, we filtered out all volumes lower than \textcolor{black}{ $V^+ = 30^{3}$~\cite{Lozano2012},} which corresponds to \textcolor{black}{ 0.77\%} of the total volume of the channel. 

\begin{figure}[H]
\centering
\includegraphics[width=1\textwidth]{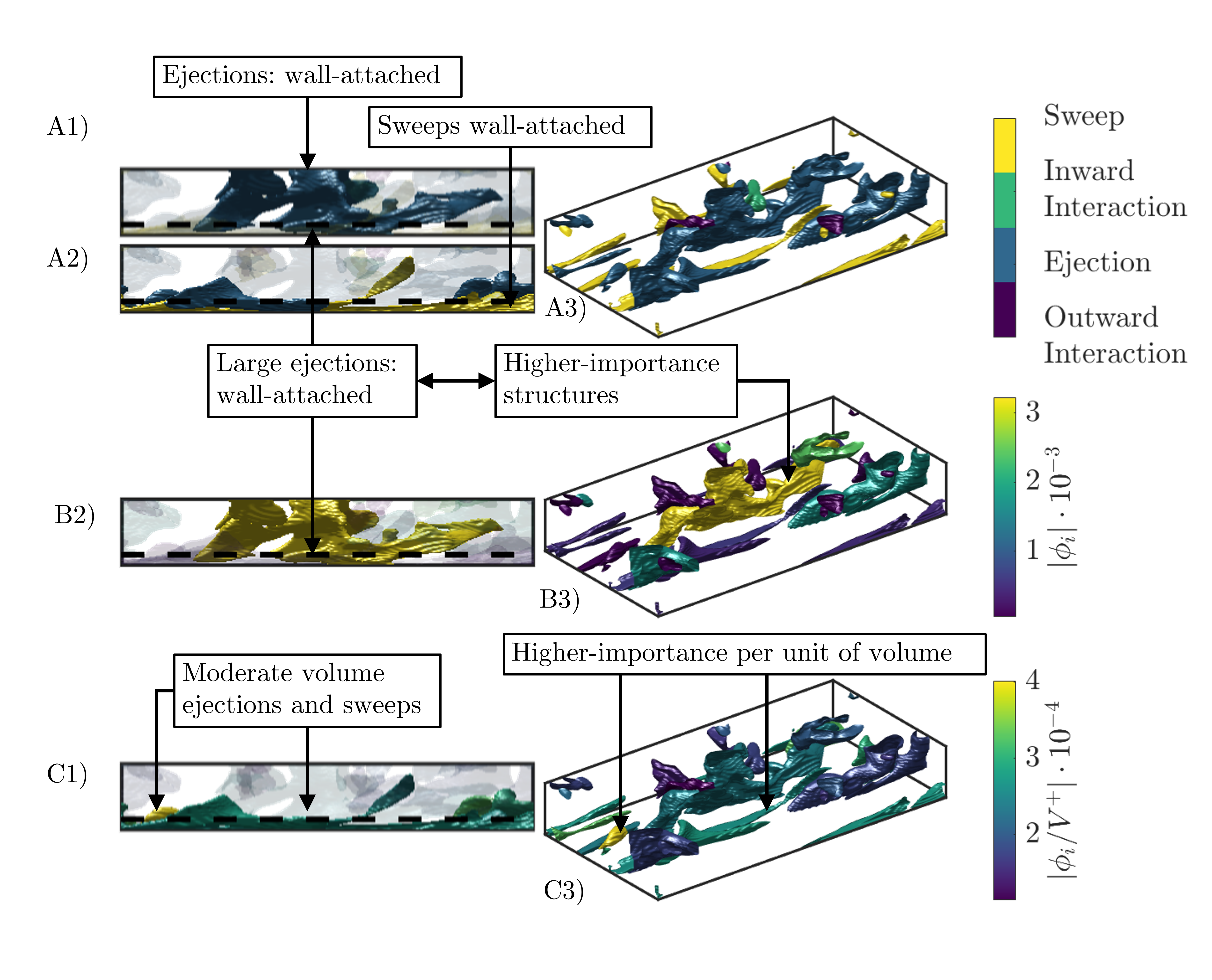}
\caption{\textcolor{black}{\bf Instantaneous visualization of the turbulent structures.} \textcolor{black}{This Figure shows (views A) the type of turbulent structure, (views B) the SHAP values and (views C) the SHAP values divided by the volume of the corresponding structures. The three-dimensional perspective is presented in images A3, B3, and C3. The side view of the turbulent channel (left) highlights the more influential structures (views A2 and B2). The most important structures per unit of volume are highlighted in views A1 and C1. Note that the highest SHAP values are obtained for large wall-attached ejections, while the moderate-size wall-attached ejections and sweeps exhibit the highest influence per unit of volume. The dashed line marks $y^+=20$, which was used in previous studies~\cite{Lozano2012} to separate wall-attached and wall-detached structures. The visualization is presented for half of the channel in all the subfigures.}}
\label{fig:shap_uv}
\end{figure}

In Figure \ref{fig:shap_uv} we can see an example of the flow and the relative importance of each structure. A wide variety of structures are present in the flow (top row, subfigures A).  Turbulence is transported in self-contained bursts composed of sweep/ejection pairs, which generate streaks as a result~\cite{Lozano2012,Jimenez2018,jimenez2022}. This idea, which was previously proposed by \citet{wal72} and \citet{lu73}, was further analysed by \citet{Lozano2012} using probability density functions of the intense Reynolds-stress structures. Using $\phi_i^e$, we can quantitatively measure the importance of every single structure. Complementing this, we can quantify the importance of every Q class. The total SHAP is presented in the top-right bar plot of Figure \ref{fig:conceptual_map}. In absolute terms, ejections are the most important events, as they represent \textcolor{black}{ 75\%} of the total SHAP score. They are followed by sweeps, with \textcolor{black}{ 22\%.} Inward and outward interactions account for the rest, as expected~\cite{Ganapathisubramani2003,Jimenez2018}. To put these percentages in perspective, approximately 60\% of the total number of structures in a turbulent channel are either sweeps or ejections, but only 25\% of them are attached to the wall~\cite{Jimenez2018}. The SHAP analysis associates the total number of sweeps and ejections with \textcolor{black}{ 97\%} of the total SHAP, \textcolor{black}{ supporting} the idea that the momentum transport relies on the self-contained bursts or ejection/sweep pairs.   Moreover, the summation of the wall-attached sweeps and ejections corresponds to a volume of approximately \textcolor{black}{ $6.9\%$} of the total, and a contribution of around \textcolor{black}{ $37.3\%$} of the Reynolds-shear-stress profile. \textcolor{black}{ These numbers are similar, although lower than the ones reported by \citet{Lozano2012}, $8\%$ and $60\%$ respectively,} a fact that may be explained by the Reynolds number considered here, which is an order of magnitude lower than those in Ref.~\cite{Lozano2012}. \textcolor{black}{ For example, Deshpande and Marusic~\cite{Deshpande_Marusic} analyzed experimental datasets across a decade of $Re_{\tau}$ and found that the relative contribution from sweeps to the total Reynolds shear stresses increases at a much faster rate than from ejections with increasing Reynolds number.} Note that, while in the work by \citet{Lozano2012} the metric used to assess the importance of the various Q events is their respective contribution to the Reynolds-shear-stress profile, in this study we consider the SHAP value instead. Interestingly, based on the SHAP metric the importance of wall-attached sweeps and ejections is \textcolor{black}{ 95.7\%} despite their low combined volume \textcolor{black}{ (19.1\% of the total)}, a fact that suggests that it may be a robust and objective metric to evaluate the importance of various coherent structures. As shown in Figure \ref{fig:conceptual_map}, ejections are the largest structures. The size of ejections can also be appreciated in the slices A2) and B2) of Figure \ref{fig:shap_uv}. Using the SHAP \textcolor{black}{ per unit volume defined in equation~(\ref{eq:density})}, the  contribution of each type of structure is modified, and \textcolor{black}{ sweeps ($44\%$) become the most influential structures per unit volume, while ejections ($42\%$) are the second most important ones.} Due to their small volume, the inward and outward interactions have a larger impact than in absolute terms, with approximately \textcolor{black}{ 14\%} of the total SHAP score. However, this is still small compared with \textcolor{black}{ the impacts of} ejections and sweeps. 

\begin{figure}[H]
\centering
\includegraphics[width=.39\textwidth]{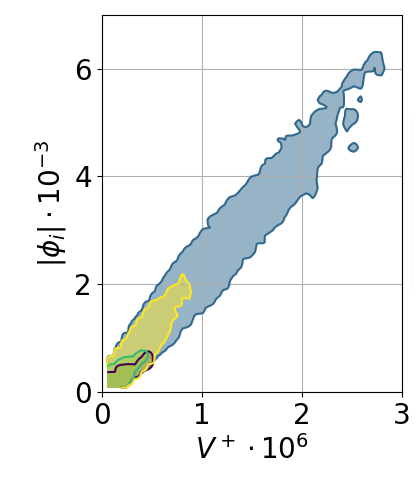}
\includegraphics[width=.59\textwidth]{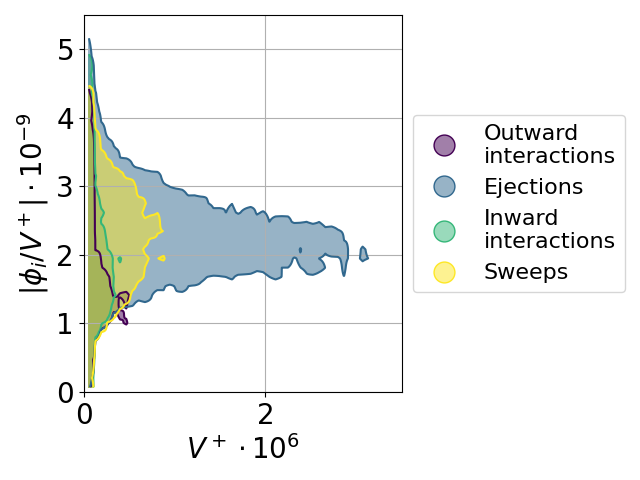}
\caption{\textcolor{black}{\bf  Magnitude of the SHAP values (left) and SHAP values per unit of volume (right) of the structures for different turbulent events as a function of their volume, expressed in inner units.} \textcolor{black}{The SHAP values determine the importance of the various turbulent structures, {\it i.e.} the most relevant structures exhibit a higher magnitude of the SHAP value. High-volume ejections are the most important structures for the predictions, while wall-detached structures, mainly medium-size ejections, exhibit a high importance per volume. These structures are often associated with a low Reynolds stress and therefore their importance is typically not identified by the methods based on contribution to the Reynolds-stress profile. Note that the distinction between wall-attached and wall-detached structures is presented in the Supplementary Material.}}
\label{fig:fig_volshapevent_back}
\end{figure}

\textcolor{black}{ Two} different families of structures are observed~\cite{del06}: wall attached, in which the lowest point is located at $y^+ < 20$ (Figure~\ref{fig:shap_uv}, A2), and wall detached, in which the lowest point is located at $y^+ \ge 20$ (Figure~\ref{fig:shap_uv}, small structures in A1). As stated by \citet{Lozano2012} and \citet{Jimenez2018}, the most important structures are the large ejections attached to the wall as they transport most of the Reynolds stress. To further analyze this situation, the SHAP value of the structures has been represented as a function of their volume, \textcolor{black}{ see} Figure~\ref{fig:fig_volshapevent_back}~(left). Wall-attached ejections are confirmed as the most important structures, and sweeps have an undoubtedly smaller value. This asymmetry between sweeps and ejections has been known since the work of \citet{nakagawa1977} and has also been discussed by many authors, see Ref.~\cite{Jimenez2018}. \citet{Lozano2012} estimated that the Reynolds stress associated with the sweeps is weaker than that of the ejections.  In Figure~\ref{fig:fig_volshapevent_back} it is shown that the wall-attached ejections are also the most influential structures per unit volume, a conclusion in agreement with the work of \citet{jimenez2022}. Note that wall-attached structures are associated \textcolor{black}{ with} energy production while the wall-detached ones are related to dissipation, and the work by \citet{jimenez2022} focused on the former. However, this figure evidences the presence of important wall-detached ejections per unit of volume.  This can be visualized in Figure~\ref{fig:shap_uv} A1), where the most influential structures per volume are shown with solid colors and in Figure \ref{fig:fig_volshapevent_back} (right). Additionally, note the presence of high-importance-per-volume inward interactions. These structures are of reduced volume and \textcolor{black}{ carry} a low Reynolds-stress magnitude. It is important to note that the relevance of these small structures was not identified by the traditional methodologies focused on the contribution to the Reynolds stress~\cite{Jimenez2018,Lozano2012}. Finally, note that the large ejections located along the streamwise direction are not the most important per unit of volume, being the moderate-size wall-attached ejections, small-size ejections and some inward interactions the most relevant, see Figure~\ref{fig:shap_uv} C1) \textcolor{black}{ and Figure~\ref{fig:fig_volshapevent_back}~(right).}

\begin{figure}[H]
\centering
\includegraphics[width=.39\textwidth]{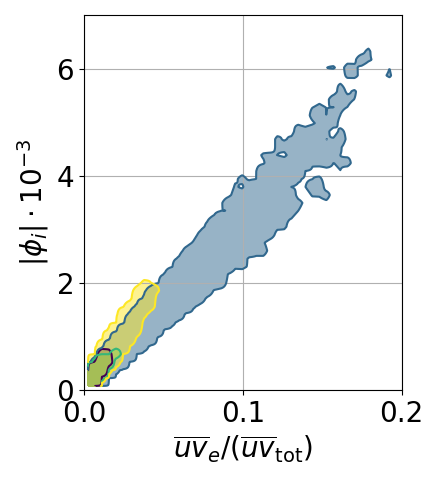}
\includegraphics[width=.59\textwidth]{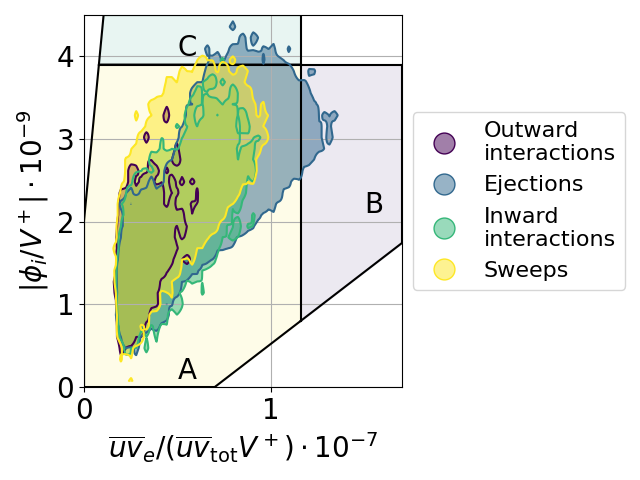}
\caption{\textcolor{black}{\bf Magnitude of the SHAP values as a function of the fractional contribution to the total Reynolds shear stress (left) and same quantities scaled with the structure volume (right).} \textcolor{black}{The left figure shows a clear relationship between the SHAP values and the contribution to the Reynolds stress; this correlation is connected with the structure size. The right panel shows some differences, since the highest SHAP values are obtained for structures which do not have the highest fractional contribution to the total Reynolds shear stress. In this panel we highlight different regions: region A (yellow) is a band containing most of the structures, region B (purple) contains the structures with the highest fractional contribution to the total Reynolds shear stress and region C (blue) contains the structures with the highest SHAP per volume. Note that the distinction between wall-attached and wall-detached structures is presented in the Supplementary Material.}}
\label{fig:fig_uvshap}
\end{figure}

As mentioned above, in this work we use the SHAP score to assess the importance of the various Q events in the flow, as opposed to calculating their contribution to the total Reynolds-shear-stress profile $\overline{uv}_{{\rm tot}}$ as previously done in the literature~\cite{lozano2014thesis}. Here we study the differences between both methods by computing the total Reynolds stress associated with each structure $\overline{uv}_e$, defined as:

\begin{equation}
    \overline{uv}_e = \int_e u(x,y,z)v(x,y,z) {\rm d} V,
\end{equation}

where the integration is done for every structure $e$ and $V$ is the volume of the structure. Without taking into account the volume, see Figure~\ref{fig:fig_uvshap}~(left), there exists a clear correlation between $\overline{uv}_e/\overline{uv}_{\rm tot}$ and the SHAP values. The larger a structure is, the more Reynolds stress the structure carries and the larger its SHAP is. However, when scaling the SHAP and $\overline{uv}_e/\overline{uv}_{\rm tot}$ distributions by the volume, the results are very different as shown in Figure~\ref{fig:fig_uvshap}~(right):  the correlation between these two quantities \textcolor{black}{ essentially disappears}. Most of the structures are located in region A, which \textcolor{black}{ exhibits a broad range of SHAP values for the same} $\overline{uv}_e/\overline{uv}_{\rm tot}$ per unit volume. \textcolor{black}{ Interestingly, the} most important structures (located in region C) are not \textcolor{black}{ necessarily the ones with} the maximum Reynolds stress per volume (found in region B). Note that the structures with the maximum specific shear stress are of low volume, while the large structures are associated with a medium specific shear stress. \textcolor{black}{ This illustrates the fact that the SHAP value is an objective quantity to measure the importance of the various coherent structures, regardless of their physical connection with the Reynolds shear stress. In fact,} the SHAP score can \textcolor{black}{ effectively} detect relatively small structures with the highest impact on the momentum transport \textcolor{black}{ per unit volume}. 

\textcolor{black}{ We further analyze the characteristics of the coherent structures related to their SHAP values in the Supplementary Material, where we study the geometrical properties of the most important structures.} \textcolor{black}{ Our results show that the streamwise-elongated wall-attached ejections, which satisfy $3<\Delta x/\Delta y<6$, exhibit the highest importance for the predictions. Note that $\Delta x$ and $\Delta y$ are the streamwise and wall-normal lengths of the box circumscribing the structure. The second most relevant structures are streamwise-elongated wall-attached sweeps.}

\textcolor{black}{ To summarize, the explainable-AI} methodology \textcolor{black}{ provides an objective measure of importance for the various coherent structures identified in the channel. Leveraging the potential of this method, we find that the} low-volume wall-attached ejections are the most influential ones \textcolor{black}{ per unit volume}, followed by low-volume wall-detached ejections and some outlier low-volume inward \textcolor{black}{ and outward interactions.} 

\section*{\textcolor{black}{ Application to an experimental dataset at higher Reynolds number}}

\textcolor{black}{ As mentioned in the Introduction, one of the advantages of this framework is the fact that it can be applied  to environments where not so much data is available but the hierarchy of energy-containing scales is broader, for instance in the context of experiments. This is a crucial point, because the structures identified as most important in the DNS at lower Reynolds number may not be the same as the most important ones in a high-Reynolds-number experiment, and also it might not be possible to measure them with sufficient fidelity in a real experimental setup. Therefore, we applied the SHAP framework to a real experimental database obtained by Lee et al.~\cite{lee_et_al_tsfp}. In this database, a turbulent boundary layer develops on a flat plate which is towed through a water tank, and measurements are carried out on a two-dimensional (2D) vertical plane using time-resolved particle-image velocimetry (PIV)~\cite{Elsinga_Marusic}. A total of 5,978 2D instantaneous flow fields of $u$ and $v$ with a spacing $\Delta t_f^+=1.5$ are analyzed, at a friction Reynolds number of $Re_{\tau}=1,377$. Note that although this Reynolds number is within the reach of what is possible with DNS, we want to illustrate the differences observed at various Reynolds numbers and using datasets with limited data availability ({\it i.e.} in experiments). We train a U-net similar to that described in the Methods section but adapted to 2D data, and we obtain relative errors of around $2\%$ in $u$ and $v$, which again leads to a very good representation of the original flow. We conduct a percolation analysis leading to a value of $H=0.54$ maximizing the number of structures, which differs from the value used in the DNS due to the 2D nature of the identified structures.}

\begin{figure}
\centering
\includegraphics[width=.39\textwidth]{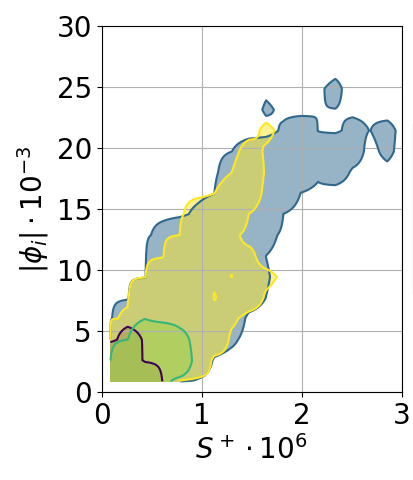}
\includegraphics[width=.59\textwidth]{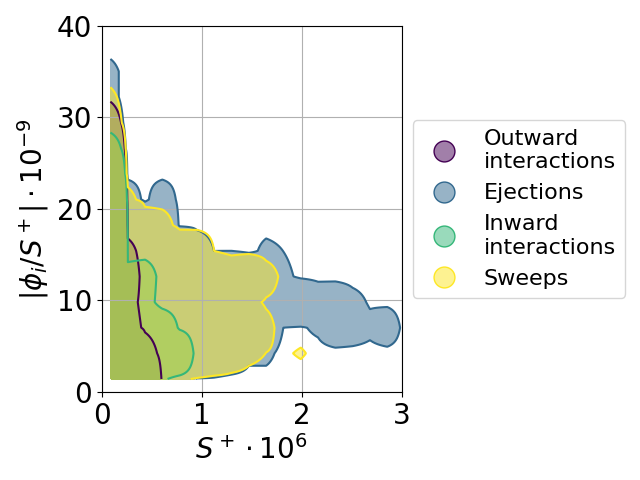}
\caption{\textcolor{black}{\bf Magnitude of the SHAP values (left) and SHAP values per unit of surface (right) of the structures in the experimental dataset~\cite{lee_et_al_tsfp} as a function of their surface, expressed in inner units.} \textcolor{black}{Note the similarity with the numerical results in Figure~\ref{fig:fig_volshapevent_back}, where the larger presence of sweeps is due to the higher Reynolds number.}} 
\label{fig:exp1}
\end{figure}

\textcolor{black}{ Computing the SHAP values leads to a distribution similar to that of Figure~\ref{fig:conceptual_map}, although with a somewhat larger importance of sweep events compared with ejections (which are still the most important ones). This increase in the importance of sweeps at higher $Re$ has also been reported in the literature~\cite{Deshpande_Marusic}. In Figure~\ref{fig:exp1}~(left) we show the magnitude of the SHAP values from all the identified structures as a function of the inner-scaled structure surface $S^+$, which shows a correlation between the structures with highest SHAP and structure surface. Note that, since the experimental data does not contain well-resolved data below $y^+\simeq 40$, we do not differentiate between wall-attached and detached structures, but the trend is consistent with that of the simulations, with the large ejections being the most important ones. In fact, when scaling the SHAP value by the structure surface, as shown in Figure~\ref{fig:exp1}~(right), we obtain essentially the same behavior as that identified in the DNS, with the largest structures reducing their SHAP per unit surface, and the medium-size ejections (and some sweeps) having the highest importance per unit surface. Also in the experimental case we identify some small inward/outward interactions which become very important when scaled with their surface.} 

\begin{figure}
\centering
\includegraphics[width=.39\textwidth]{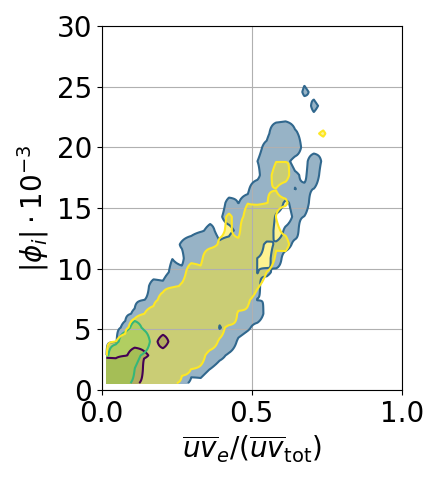}
\includegraphics[width=.59\textwidth]{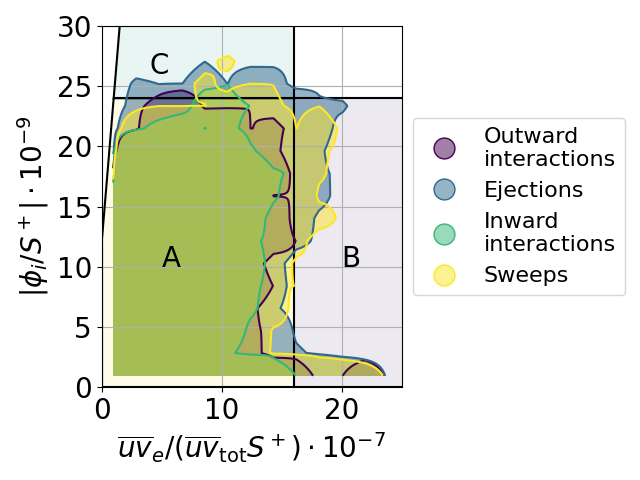}
\caption{\textcolor{black}{\bf Magnitude of the SHAP values as a function of the fractional contribution to the total Reynolds shear stress (left) and same quantities scaled with the structure surface (right) in the experimental dataset~\cite{lee_et_al_tsfp}.} \textcolor{black}{In the right panel we highlight different regions: region A (yellow) is a band containing most of the structures, region B (purple) contains the structures with the highest fractional contribution to the total Reynolds shear stress and region C (blue) contains the structures with the highest SHAP per volume. Note the similarity with the numerical results in Figure~\ref{fig:fig_uvshap}, where the larger presence of sweeps is due to the higher Reynolds number.}} 
\label{fig:exp2}
\end{figure}

\textcolor{black}{ The comparison between SHAP values and contribution to the Reynolds shear stress is also performed for the experimental data, and is shown in Figure~\ref{fig:exp2}. In Figure~\ref{fig:exp2}~(left) we show that in the experiment there is also a good correlation between the magnitude of the SHAP value and the contribution to the Reynolds shear stress. Interestingly, Figure~\ref{fig:exp2}~(right), where we divide both by the structure surface, also shows the 3 regions detected in the DNS data. In region A, there is a broad spread of Reynolds shear stress contributions for the same SHAP value; in region B we identify the structures with the highest Reynolds shear stress, which are not necessarily the ones with the highest SHAP; and finally, in region C we find the structures with the highest SHAP per unit surface, which again exhibit a quite broad range of values of the Reynolds shear stress. Note that the mean spanwise size of the structures in the DNS is about 11.6 viscous units; thus, if one would calculate a surrogate of the structure volume in the experiment using this value, a good qualitative agreement between Figures~\ref{fig:exp2}~(right) and \ref{fig:fig_uvshap}~(right) would be obtained, which is reassuring. This figure shows, also for an experimental dataset at a higher Reynolds number, that the present XAI framework can identify the most important structures in the flow in a more objective way than the calculation of their contribution to the Reynolds shear stress, a fact that can have important implications in experimental campaigns at very high Reynolds numbers beyond the reach of DNS. }

\begin{figure}
\centering
\includegraphics[width=0.99\textwidth]{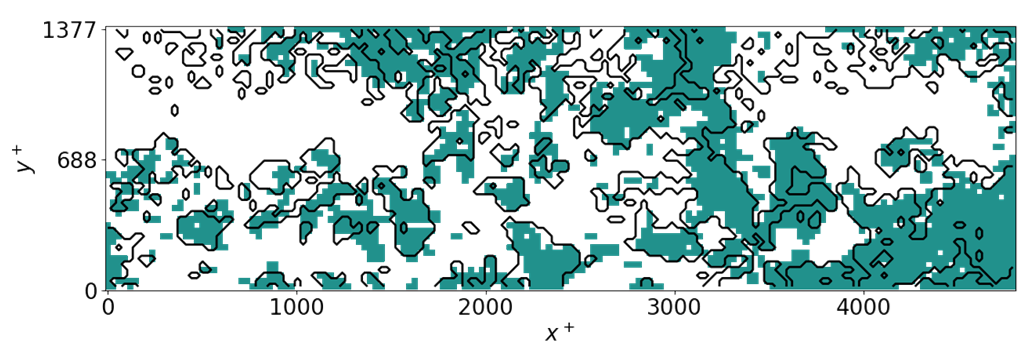}
\caption{\textcolor{black}{\bf Instantaneous flow field from the experimental dataset~\cite{lee_et_al_tsfp}, showing the difference between Q events and structures identified by means of SHAP.} \textcolor{black}{In this figure the green color denotes structures identified by means of the SHAP method, whereas the black contours indicate the Q events, both identified in terms of their corresponding criteria applied to each point in the domain and percolation analysis. The difference between both types of structures is around $30\%$, measured as the fraction of points belonging to Q events and not to SHAP structures and vice versa.}} 
\label{fig:SHAP_Q_comparison}
\end{figure}

\textcolor{black}{ To further illustrate the potential of the present methodology to identify regions of importance of the flow, we carry out an additional SHAP analysis on the experimental dataset, although this time the input features are not the Q events, but rather each individual point of the measured fields. Doing so, instead of classifying the Q events by importance we effectively define completely new structures based on their SHAP values, and assess how different they are from the Q events.} \textcolor{black}{ The increased number of input features motivated the usage of a slightly different SHAP algorithm, namely gradient SHAP~\cite{lundberg2017}, which is computationally more efficient.} \textcolor{black}{ The main difference with respect to the kernel-SHAP technique discussed in the Methods section is the fact that it yields different SHAP values for each of the velocity-fluctuation components, {\it i.e.} $u$ and $v$, which we will denote as $q_u$ and $q_v$. In order to define the 2D structures based on the SHAP criterion, we consider the following equation:}
\begin{equation}\label{SHAP_percolation}
\textcolor{black}{\sqrt{ q_u^2(x,y,t) + q_v^2(x,y,t)}  > H_q \sqrt{ q_u^{\prime 2} (x,y) + q_v^{\prime 2} (x,y)},}
\end{equation} 
\textcolor{black}{ where $q_u^{\prime}$ and $q_v^{\prime}$ are the rms of the SHAP values corresponding to the $u$ and $v$ fluctuations calculated based on the whole experimental dataset. Note that here we use the norm of the SHAP score to account for the contributions in both directions. A percolation analysis is carried out on equation~(\ref{SHAP_percolation}), revealing that the value $H_q=0.55$ maximizes the number of identified coherent structures. A comparison between the structures identified by means of the SHAP method and the Q events for a sample instantaneous snapshot is shown in Figure~\ref{fig:SHAP_Q_comparison}. This figure suggests that, despite the fact that both types of structures share some similarities, they are indeed different, a result in agreement with the previous discussions (namely that the most important structures are not necessarily the ones with the highest contribution to the Reynolds shear stress). In fact, analyzing the overlap between both types of structures for the entire database, we obtain that they only overlap by around $70\%$, and therefore there are important differences between both types of structures which will be analyzed in further detail in future work.}

\section*{Discussion and conclusions}\label{sec:conclusion}

This article uses XAI to \textcolor{black}{ quantify} the importance \textcolor{black}{ of the various types of} intense Reynolds-stress structures. Ejections and sweeps are shown to be the most relevant structures for fully-developed turbulence in a channel. A similar conclusion was reported by \citet{Lozano2012} by considering the contribution of the Q events to the Reynolds shear-stress profile as the metric indicating structure importance. \textcolor{black}{ However, when studying the importance of the various Q events per unit volume, we find that the structures with the largest contribution to the Reynolds shear stress are not the ones with the largest importance score.} The SHAP value considered here relies on the contribution of the structures to predict \textcolor{black}{ an instantaneous flow field several viscous units in the future,} a fact that leads to a more objective assessment of structure \textcolor{black}{ importance.} The structures \textcolor{black}{ were first} extracted from the simulation of a turbulent channel and used to segment the computational domain. Then, their importance on the flow was calculated by measuring their contribution to the \textcolor{black}{ U-net} prediction. \textcolor{black}{ An experimental dataset consisting of time-resolved 2D PIV measurements was also used to test the present framework.}

Ejections and sweeps are highly important in generating turbulent self-contained bursts \cite{Jimenez2018,jimenez2022} and are associated with turbulence production. The higher-volume structures are the most influential when assessing the global prediction, \textcolor{black}{ and they} correspond to wall-attached ejections extending throughout the channel, transporting a substantial fraction of the total Reynolds stress~\cite{Lozano2014}. These ejections exhibit the largest modulus of the SHAP value, meaning that their presence is essential for the correct prediction by the \textcolor{black}{ U-net}. However, different trends are obtained when analyzing structure influence relative to the volume. In this case, the most influential structures per unit volume are \textcolor{black}{ relatively} smaller wall-attached ejections, \textcolor{black}{ although} this methodology evidences the local importance of small-size wall-detached structures \textcolor{black}{ as well. In low-Reynolds-number flows, these} structures are mainly ejections, although some inward interactions have shown local importance; \textcolor{black}{ the relative importance of sweeps is found to increase at higher Reynolds number, which is consistent with the trends noted in previous experimental observations~\cite{Deshpande_Marusic}.} These results support the idea of using SHAP values for analyzing turbulent structures, thus enabling the extraction of deeper knowledge on the turbulent flow.

Relative to the shape of the structures, the most influential structures per unit volume exhibit larger aspect ratios in the streamwise direction than in the spanwise direction. In addition, the wall-normal length is 3--6 times smaller for the structures with \textcolor{black}{ the} highest SHAP and SHAP per unit volume. Furthermore, the structures with higher specific importance are contained in the $xy$ plane, being the aspect ratio in the $z$ direction lower. 

The framework presented here has enabled, in a purely data-driven manner, to confirm and expand some of the basic knowledge of wall-bounded turbulence available in the literature~\cite{Jimenez2018,Lozano2012}. \textcolor{black}{ Furthermore, when applying the SHAP framework to each point in the domain individually, we found that there is around 30\% mismatch between the Q events and the structures with the highest importance score. This conclusion will be further investigated in future work, with the aim of } shedding light on the fundamental phenomena of wall-bounded turbulence. In terms of turbulence modeling, a similar approach may be taken to first quantify and subsequently understand the significance of coherent structures and of dynamical processes such as vortex stretching and strain amplification in data-driven subgrid-scale representations, for instance, when based on invariants of the velocity-gradient tensor. 

 Furthermore, the present methodology may help to gain tremendous insight into the basic mechanisms of wall-bounded turbulence. As indicated above, turbulent flows are ubiquitous in a wide range of problems of great industrial and environmental interest, such \textcolor{black}{ as} combustion, aerodynamics, energy generation, transportation and the current climate emergency. Obtaining detailed knowledge on the building blocks of turbulence will be instrumental to be able to control these flows, thus obtaining great gains in all these important applications. Note however that, in order to use the SHAP framework presented here, it is important to obtain a detailed representation of the coherent structures in the flow. One approach is to perform DNS, which is progressively enabling detailed simulations of complex flows, such as turbulent wings, where coherent structures can be identified~\cite{atzori_wing}. \textcolor{black}{ Nevertheless}, the very high computational cost of DNS~\cite{choi_mesh} precludes the application of this method for full-scale applications, at least at the moment. However, rapid development of computational facilities, particularly in the context of graphics-processing-unit (GPU)-accelerated architectures, may enable very detailed simulations at very high Reynolds numbers in the next years. Furthermore, experimental work in fluid mechanics is progressively benefiting more from machine learning~\cite{ai4exp}, and it might be possible to obtain high-fidelity flow representations at much higher Reynolds numbers, thus enabling the usage of SHAP frameworks in more practical flow cases. \textcolor{black}{ Note that, based on this work, the structures identified by means of SHAP in a high-fidelity simulation at lower Reynolds number might not be easy to measure in an experiment at high Reynolds number with lower resolution. Therefore, one advantage of the present framework is its applicability directly to the experimental data, thus enabling a more direct way to detect (and control) the most important structures. Flow} control by deep \textcolor{black}{ reinforcement} learning (DRL) is already leading to impressive drag-reduction rates in turbulent flows~\cite{guastoni_drl}, and being able to leverage DRL to control the most important flow structures identified via SHAP may constitute a novel paradigm in terms of flow control, increasing the potential for reducing energy consumption in transportation.

\section*{Methods}\label{sec:method}

\subsection*{Numerical simulations and flow case under study} \label{sec:case}

The \textcolor{black}{ U-net} was trained using \textcolor{black}{ 6,000} instantaneous velocity fields obtained through DNS, \textcolor{black}{ with a spacing of $\Delta t^+_f=5$. Note that analyses carried out with $\Delta t^+_f=1$ and 2 led to results consistent with those with $\Delta t^+_f=5$.} The simulations are calculated in a box with periodic boundaries confined between two parallel plates and driven by an imposed pressure gradient. The employed code is LISO~\cite{llu21a}, which has been used to run some of the largest simulations of wall-bounded turbulence~\cite{hoy22}. The convergence of the turbulence statistics was assessed based on the criterion of linear total shear~\cite{Vinuesa2016b}. The data obtained with LISO has been extensively validated against experimental and other numerical studies~\cite{hoy06,hoy08} and is broadly used~\cite{mon21,spa21,pirozzoli2022dns}. 

\subsection*{Deep-neural-network architecture and prediction}

A \textcolor{black}{ U-net architecture~\cite{ronneberger2015}} is used for predicting the velocity field. \textcolor{black}{ This architecture efficiently exploits spatial correlations in the data, and further develops the work by} \citet{Schmekel2022}. Note that \textcolor{black}{ U-nets} and other computer-vision architectures have been successfully used in the context of turbulent-flow predictions~\cite{guastoni2021,guemes2021,lim1,lim2,lim3}. The convolution operation is described by Equation (\ref{eq:convneur}), where \textcolor{black}{ $f_i$} is the input three-dimensional (3D) tensor, $h$ the filter, $G$ the output, and $m$, $n$ and $p$ the indices of the output tensor:
\begin{equation}\label{eq:convneur}
G(m,n,p) = (f_i*h)(m,n,p) = \sum_i \sum_j \sum_k h(i,j,k)f_i(m-i,n-j,p-k).
\end{equation}

\begin{figure}[H]
\centering
\includegraphics[width=1\textwidth]{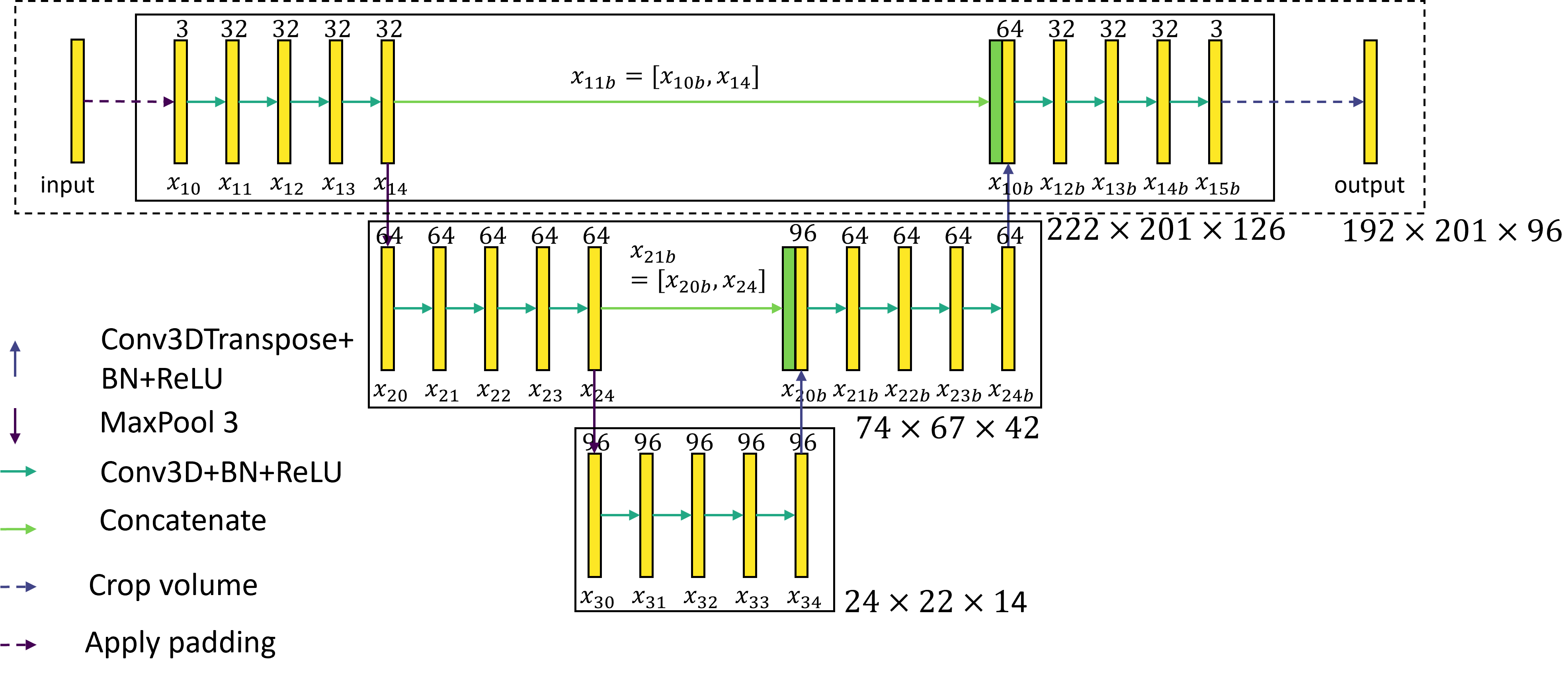}
\caption{\textcolor{black}{\bf Schematic representation of the U-net architecture.} \textcolor{black}{In this representation, $x_{ac}$ denotes each layer, where $a$ indicates the level, $c$ the number of layer and the letter $b$ is associated with layers after the bottleneck (located at the bottom, {\it i.e.} level 3). The number above each layer is the number of filters, and the size of the field in grid points is also indicated for each of the 3 levels. Note that the pooling and upsampling operations reduce and increase the size of the fields, respectively, and `BN' stands for batch normalization.}}
\label{fig:ske_archi}
\end{figure}

\textcolor{black}{ The U-Net comprises 20 layers of 3D convolutional-neural-network (CNN) blocks, 2 max poolings, 2 transposed 3D CNN blocks, and 2 concatenations in the layers~\citep{Kaiming2016}. Padding is added to the boundaries in the spanwise and streamwise directions of the channel reproducing the periodic information of the flow field. As a consequence of the padding, the original size of the volume, $192\times201\times96$ points (in $x$, $y$ and $z$ respectively), is increased to $222\times201\times126$ points. Each convolution comprises 32 filters for the initial size of the field. The number of filters is increased to 64 and 96 after the first and second max pooling, a fact that reduces the size of the fields to $74\times 67\times42$ and $24 \times 22\times14$ respectively. In addition, each filter has a size of $3\times 3\times 3$. Rectified-linear-unit (ReLU) activation functions are used to avoid the vanishing-gradients problem~\citep{tan2019}. The network setup is selected to obtain an adequate accuracy with a reasonable computational cost. The network, which is shown in Figure~\ref{fig:ske_archi}, } uses a total of \textcolor{black}{ approximately 2 million parameters (where 99.9\% of them are trainable).} Here, \textcolor{black}{ 67\%} of the flow fields are used for the training-and-validation process (out of which 80\% are used for training and 20\% for validation). For this process, an RMSprop optimizer is used \cite{zou2019}. The remaining \textcolor{black}{ 33\%} of fields are reserved for testing and explainability analysis, and are not seen by the network during training. The training process is concluded when the mean-square-error-based loss function is lower than \textcolor{black}{ $5 \times 10^{-4}$,} corresponding to $1.5\times 10^4$ epochs, where all training data is
used once in a single epoch.

\subsection*{Explainability of the neural network}

Despite the excellent results achieved with deep learning, the relationships between inputs and outputs are complex, and it is, in general, challenging to explain the predictions based on a particular input field.  {It is common to use linear models, based on additive feature-attribution methods, such as the one shown in equation (\ref{eq:linregres}), for interpretation.}
 
 {\begin{equation}\label{eq:linregres}
g(q') = \phi_0 + \sum_{j=1}^{\lvert Q \lvert} \phi_j q_j'.
\end{equation}}  

 {In equation (\ref{eq:linregres}), $g(q')$ is an approximation of the error between the U-net and the ground truth, $q_j'$ are binary variables representing the presence or absence of each feature (the coherent structures in our case), and $\phi_j$ are the SHAP values. Note that $\phi_0$ would be the model output when all the features are removed. This work uses the kernel-SHAP algorithm~\cite{lundberg2017} in order to calculate the importance of each turbulent structure in the model prediction. Kernel-SHAP is an additive feature-attribution method which calculates $g(q')$ as a sum of the SHAP values associated with all the input features. This algorithm is based on the combination of two different techniques: LIME~\cite{Ribeiro2016} and Shapley values~\cite{Lipovetsky}.}

 {LIME~\cite{Ribeiro2016} interprets the individual predictions of the model by approximating them locally, as in equation~(\ref{eq:lime}). This local explanation adheres to the additive feature-attribution method defined in equation~(\ref{eq:linregres}). The contribution of each feature is calculated by minimizing a loss function, $\mathcal{L}$, which depends on the explanation model, $g$, the error between the original model and the ground truth, $f$, and a local kernel, $\pi_x$. Additionally, the complexity of the model is penalized by including the term $\Omega(g)$.}
 {\begin{equation}\label{eq:lime}
\xi = \arg \min_{g \in \mathcal{G}}{\mathcal{L} (f,g,\pi_x)+\Omega(g)}.
\end{equation}}

 {In order to produce a unique solution, the LIME methodology requires to satisfy the local-accuracy, missingness and consistency properties~\cite{lundberg2017}. However, equation~(\ref{eq:lime}) cannot ensure the previous properties for a heuristic definition of its parameters.}  {These properties are satisfied by the classical Shapley-value estimation methods~\cite{Lipovetsky}. These values are derived from an axiomatic approach, ensuring that they satisfy the unique-solution requirements. In the present context, the Shapley values quantify the marginal contribution of a particular structure $i$ (this will be denoted as feature) to the error in the model $f$ when included in a particular group of structures $s$ (which will be denoted as coalition). These values are calculated as follows:}
 {\begin{equation}\label{eq:shapleyvalues}
    \phi_i(Q,f) = \sum_{s \subseteq Q\backslash \lbrace i \rbrace} \frac{ \lvert s \lvert !( \lvert Q \lvert - \lvert s \lvert -1)!}{ \lvert Q \lvert !}\left ( f(s \cup \lbrace i \rbrace)-f(s)\right ),
\end{equation}}
 {\noindent where $Q$ is the set containing all the structures in a particular field, $ \lvert Q \lvert $ is the total number of structures and $ \lvert s \lvert $ is the total number of possible coalitions not containing structure $i$. The expression $ \lvert s \lvert !( \lvert Q \lvert - \lvert s \lvert -1)!$ represents all the possible combinations of structures where $s$  is the coalition before evaluating $i$ and $\lvert s \lvert !( \lvert Q \lvert - \lvert s \lvert -1)! / \lvert Q \lvert !$ the probability that the structure $i$ is included in the model after the coalition of structures $s$.}

 {Although the Shapley values can quantify the marginal contribution of each structure, their exact computation is challenging. This problem is solved by approximating their solution using the so-called kernel-SHAP methodology~\cite{lundberg2017} mentioned above. As previously stated, the methodology  combines the solution of LIME~\cite{Ribeiro2016} with Shapley values~\cite{Lipovetsky}. Equations~(\ref{eq:complexitypenalty})--(\ref{eq:kernelShap}) show the values of the regularization term, the loss function and the weighting kernel that recover the definition of the Shapley values. }

 {\begin{equation}\label{eq:complexitypenalty}
    \Omega(g) = 0,
\end{equation}}
 {\begin{equation}\label{eq:lossshap}
\mathcal{L}(f,g,\pi_x) = \sum_{q'\in Q}\left[f(h_x(q'))-g(q')\right]^2\pi_x(q'),
\end{equation}}
 {\begin{equation}\label{eq:kernelShap}
\pi_x(q') = \frac{\lvert Q \lvert-1}{\left(\begin{array}{c}
\lvert Q \lvert\\
\lvert q' \lvert \\
\end{array}\right) \lvert q' \lvert \left(\lvert Q \lvert - \lvert q' \lvert\right)}.
\end{equation}}
 {\noindent Here $ \lvert q' \lvert $ is the number of nonzero structures and $h_x$ a mask function which converts the binary space of $q'$ into the space of the input of the model. Then, the LIME equation is solved using linear regression, with very low errors of the order of $(f-g)^2 = \mathcal{O}(10^{-7})$.}

\begin{figure}[H]
\centering
\includegraphics[width=1\textwidth]{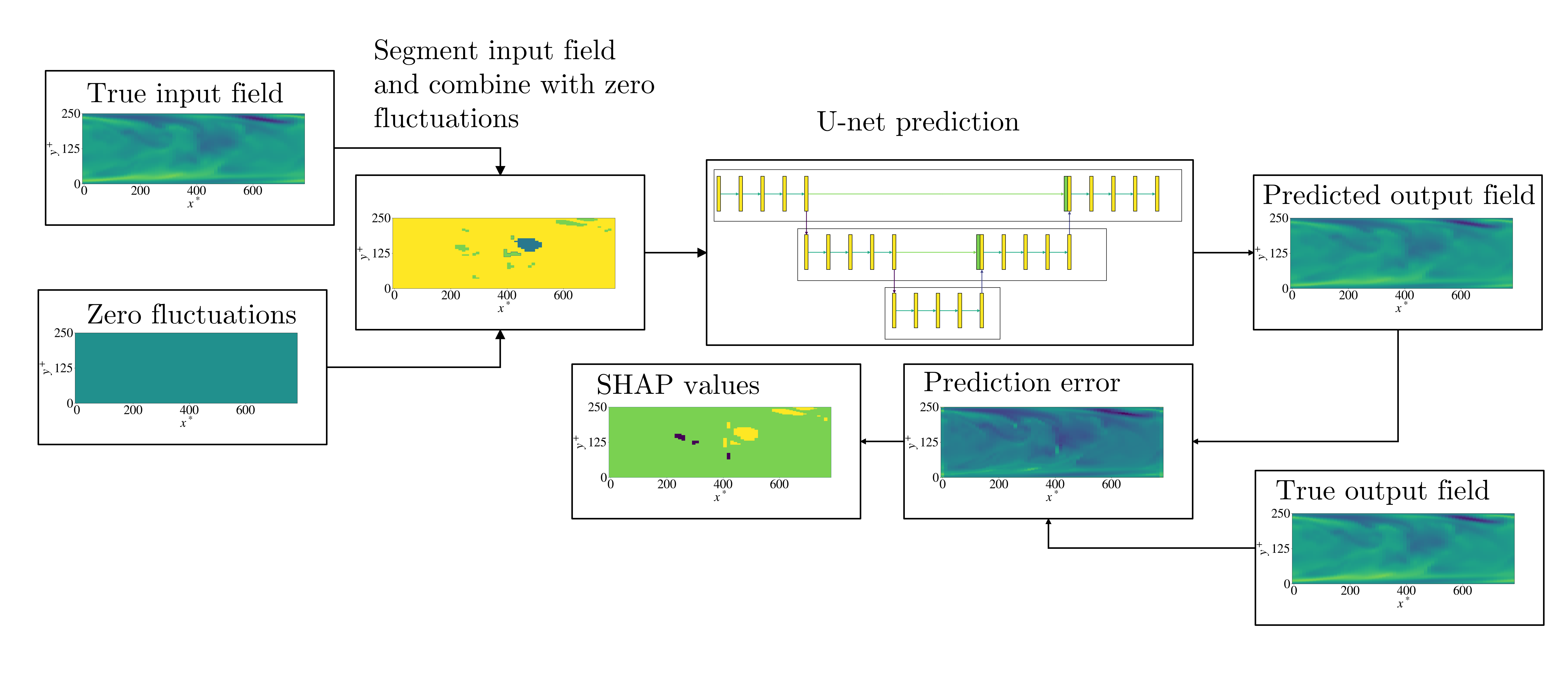}
\caption{ {\textbf{Implementation of the kernel-SHAP algorithm.}} \textcolor{black}{The kernel-SHAP algorithm can be divided into different stages. In the first one, the domain is segmented: a coalition is chosen, where blue denotes the structure being analyzed and green the rest; yellow denotes the background. Then, the results are predicted using the U-net for each structure. Finally, the error between this prediction and the ground truth is used to determine the SHAP value of each feature. In the SHAP panel, purple and yellow denote low and high values of SHAP respectively, while green denotes the background.}}
\label{fig:shap_procedure}
\end{figure}

 {In the context of wall-bounded turbulence, the importance of the Reynolds-stress structures is the focus of our analysis. Thus, a mask function is defined to map from the original space into the space of the structures. For each instantaneous field, a total of $2\lvert Q\lvert + 2048$ different coalitions is used in the kernel-SHAP method, where $\lvert Q\lvert$ is typically around 150 in the channel case. Note that these coalitions are selected randomly, as a representative sample of all the possible coalitions (which would be a computationally intractable problem). Then, the importance of each structure is calculated as its influence on the predictions, where the absent structure of each coalition is substituted by a region of zero fluctuations.  Besides this, we also considered two additional cases: one where, instead of replacing the structure with a region of zero fluctuations, we scaled down the $u$ and $v$ fluctuations such that they are 5\% below the threshold to be considered a Q event, while making $w=0$. In the second case, we scaled $u$ and $v$ as in the previous case, but left $w$ unchanged. In both cases the results and trends were qualitatively the same as when all the fluctuations are set to zero, and the main conclusions of this work were confirmed. The final SHAP of each structure is calculated as an average of the importance scores from all the coalitions where it was considered. To calculate the importance, the mean-squared error of the prediction is used as the output of the model $f$. The employed workflow is summarized in Figure~\ref{fig:shap_procedure}.}

 {The SHAP values represent, in a unique and quantitative manner, the contribution that a single coherent structure has on the mean-squared error of the prediction. The most relevant structures are those with a higher absolute value, while a low absolute value is related to the less relevant structures. It is important to note that, as stated above, to calculate the SHAP value of a particular structure many different coalitions are considered, where different combinations of structures are present in the field. This enables taking into account a wide range of interactions among structures, emulating a number of inter-structure interactions present in wall-bounded turbulent flows.}

\section*{Acknowledgments}
\label{sec:acknowledges}

The deep-learning-model training was enabled by resources provided by the National Academic Infrastructure for Supercomputing in Sweden (NAISS) at Berzelius (NSC), partially funded by the Swedish Research Council through grant agreement no. 2022-06725. This project has been partially funded by the Spanish Ministry of Science, Innovation, and University through the University Faculty Training (FPU) program with reference FPU19/02201 (AC). The data has been obtained with \textcolor{black}{ support} of grant PID2021-128676OB-I00 funded by MCIN/AEI/10.13039/ 501100011033 and by “ERDF A way of making Europe”, by the European Union (SH). RV acknowledges the financial support from ERC grant no. `2021-CoG-101043998, DEEPCONTROL'. Views and opinions expressed are however those of the author(s) only and do not necessarily reflect those of the European Union or the European Research Council. Neither the European Union nor the granting authority can be held responsible for them. \textcolor{black}{ RD acknowledges the financial support from the Melbourne Postdoctoral Fellowship of the University of  Melbourne. RD, JL, JPM, NH and IM are grateful to the ARC (Australian Research Council) for their continuous financial support.}

\section*{Author Contributions}

\textbf{Cremades, A.:}  Methodology, Software, Validation, Investigation, Writing - Original Draft, Visualization \textbf{Hoyas, S.:} Data curation, resources, Writing - Original Draft, Funding acquisition
\textcolor{black}{ Deshpande, R.:} \textcolor{black}{Methodology, Data curation, Writing - Review \& Editing}
\textbf{Quintero, P.:} Writing - Review \& Editing, Funding acquisition
\textbf{Lellep, M.} Writing - Review \& Editing 
\textcolor{black}{ Lee, J.:} \textcolor{black}{Data acquisition, Data curation, Writing - Review \& Editing}
\textcolor{black}{ Monty, J. P..:} \textcolor{black}{Data acquisition, Data curation, Writing - Review \& Editing}
\textcolor{black}{ Hutchins, N.:} \textcolor{black}{Data acquisition, Data curation, Writing - Review \& Editing}
\textbf{Linkmann, M} Writing - Review \& Editing 
\textcolor{black}{ Marusic, I.:} \textcolor{black}{Writing - Review \& Editing, Project administration, Funding acquisition}
\textbf{Vinuesa, R.:} Conceptualization, project definition, methodology, resources, \textcolor{black}{ Writing - Original Draft}, Supervision, Project administration, Funding acquisition.

\section*{Data availability}
The data and codes used to produce this study will be made available for open access as soon as the article is published.

\bibliographystyle{elsarticle-num-names}
\bibliography{cas-refs}

\end{document}